 \documentclass[9pt,prd,aps,amssymb, amsmath,tightenlines]{revtex4}
 \usepackage{graphics}
 \usepackage[pdftex]{graphicx}

 \usepackage{amsmath,amssymb,amsthm}

\begin{document}


\title{
Automated Liquidity: Market Impact, Cycles, and De-pegging Risk}

\author{ B. K. Meister}

\email{bernhard.k.meister@gmail.com}

\date{\today }

\begin{abstract}

\noindent
Three traits of  decentralized finance are studied.
First, the market impact function is derived for optimal-growth liquidity providers. For a standard random walk, the classic square-root impact is recovered. An extension is then derived to fit general fractional Ornstein-Uhlenbeck processes. These findings break with the `linearized' liquidity models used in most decentralized exchanges.
   
\noindent
Second, a Constant Product Market Maker is viewed as a multi-phase Carnot engine, where one phase matches the exchange of tokens by a liquidity taker, and another the change of pool size by a liquidity provider.

\noindent
Third, stablecoin de-pegging   is a form of catastrophe risk. 
By using growth optimization, default odds are linked to the cost of catastrophe bonds. De-pegging insurance can act  as a counterweight  and a key marketing tool  when the law forbids the payment of interest on stablecoins.

    \end{abstract}
\maketitle
   
\section{Introduction}
\vspace{-0.3cm}
\noindent
 All functioning markets are alike, but every dysfunctional market is dysfunctional in its own way, to paraphrase Tolstoi.
 By bypassing legacy frictions, cryptofinance has  traded old inefficiencies for new faults that can only be outgrown through iteration. 
 While some mismatches are in plain sight others are hidden. Three such properties and related design challenges are examined; but as William Blake remarked, `the errors of a wise man make your rule, rather than the perfections of a fool'. 
 
  \noindent
 Decentralized Exchanges (DEXs),  often in direct competition with Centralized Exchanges (CEXs), are one of the innovations of Decentralized Finance (DeFi) with tokens worth tens of billions of dollars currently committed by Liquidity Providers (LPs) and tokens worth billions of dollars  switched daily by Liquidity Takers (LTs). 

\noindent
The paper starts with re-deriving, extending and  applying  the widely-known square root market impact\cite{bouchaud2004subtle} of relevances for CEXs and DEXs, and then rephrases in terms of a thermodynamical heat engine the Constant Product Market Maker (CPMM) - a sub-category of the Constant Function Market Maker (CFMM). 
 A look at the pricing of catastrophe bonds follows, showing how it bears on stablecoin insurance, before a final wrap-up.
\noindent
\section{A Route to Market Impact}
\noindent
Market impact is of interest for centralised as well as decentralised markets. It will be shown how growth optimal behaviour of liquidity providers,  who allocate capital across multiple assets, 
 and a mean-reverting Ornstein Uhlenbeck process for mis-pricings leads  to a square root market impact. 

  \noindent
\noindent 
A compressed derivation is first given, which is then expanded.
The Ornstein-Uhlenbeck (OU) price process of the asset of interest is  
\begin{equation}
d P_t = \kappa (P_t-K) dt + \sigma dB_t, \nonumber
\end{equation}
with $  \Delta P:= \kappa (P_t-K)$, and the optimal Kelly fraction\cite{meister2011} is
\begin{equation}
f= \frac{\Delta P}{\sigma^2}. \nonumber
\end{equation}
For LPs investing across multiple assets their large capital $W$ times $f$ scales as $W f \propto \sqrt{Q}$.
As a consequence, since the total wealth $W$ of an active market maker adjusts slowly in respect to any single investment,   
$f$ becomes dynamic and has to scale with square root of $Q$, which immediately   
 leads to the square root impact relation:
\begin{equation}
\Delta P\propto \frac{ \sigma^2 }{W} \sqrt{Q}.  \nonumber
\end{equation}

\noindent 
Next, follows an extended version of the derivation directly relating $\Delta P$ to $Q$ without introducing $f$.
Market makers are assumed to maximize an objective function such as change of log utility:     
\begin{equation}
g = E[\ln(W + dX) - \ln(W)] = E\left[ \ln\left(1 + \frac{dX}{W}\right) \right], \nonumber
\end{equation}
where   $dX$ is the uncertain change in the value  of the position in the risky asset. 
Taylor expansion\footnote{Cubic terms and above, i.e. $O\Big(\big(\frac{dX}{W}\big)^3\Big)$, are excluded,  but in the continuous Ornstein-Uhlenbeck setting these terms are negligible.  } gives
\begin{equation}
g =E\left[ \frac{dX}{W} - \frac{1}{2}\left(\frac{dX}{W}\right)^2 \right] = \frac{E[dX]}{W} - \frac{1}{2}\frac{Var[dX]}{W^2}. \nonumber
\end{equation}
For a market maker with position $Q$, price change $\Delta P$, and volatility $\sigma$, we have $E[dX] = Q \Delta P$ and $Var[dX] = Q^2 \sigma^2$. 
This yields the fundamental growth equation 
\begin{equation}
g(Q) = \frac{Q \Delta P}{W} - \frac{Q^2 \sigma^2}{2W^2}. \nonumber
\end{equation}
\noindent   

\noindent 
The next step requires closer scrutiny, since a further constraint is imposed. 
For a diversified portfolio, capital $W$ scales with the square root of the position: $W = k\sqrt{Q}$. 
Unlike an investor with one risky holding, a multi-asset liquidity providers  only hedges the tail of the exposure, after averaging over a large number of positions, which scales with $\sqrt{N}$ instead of $N$. This is capital efficient.  
 In other words, when LPs observe flow in a particular asset, they will provide liquidity at a `price'.  This leads to LPs being saddled with market exposure requiring hedging  of risk factors using liquid assets. As an example, the first factor is often overall market exposure and the second factor might be the exposure of a stock to the oil price, or as an alternative the hedge could be determined through the Lyons' Signature Transform, which is computationally demanding but can be again asset light. 
 
 \noindent 
A large LP does not bet its whole capital on just one OU process. It sets aside a `slice' of balance sheet to back each trade. If a trade has a size $Q$, the firm’s risk model only charges capital at the rate of $\sqrt{Q}$ due to portfolio wide netting, and the Kelly growth optimzation logic   then sets the price gap $\Delta P$ accordingly. This creates investment consistency for individual LPs across a multitude of simultaneously available investment opportunities. Since the capital cost is measured in $\sqrt{Q}$, while  the trade size and returns  grow with $Q$,  the gap between them widens with size.
The `optimal fraction' $f$ is in effect subsidized by the firm's size. This is one reason besides technological scaling why LPs, and also multi-strategy hedge funds, which are governed by the same diversification logic,  have become behemoth.  

 \noindent 
 Substituting $W = k\sqrt{Q}$  into the growth equation gives
\begin{equation}
g(Q)= \frac{Q \Delta P}{k\sqrt{Q}} - \frac{Q^2 \sigma^2}{2(k\sqrt{Q})^2} = \frac{\Delta P}{k}\sqrt{Q} - \frac{\sigma^2 Q}{2k^2}.\nonumber
\end{equation}
The maximal growth rate   is obtained by setting the derivative of $g$ to zero,
\begin{equation}
\frac{dg}{dQ}  = \frac{\Delta P}{2k\sqrt{Q}} - \frac{\sigma^2}{2k^2} = 0,\nonumber
\end{equation}
which ties  $\Delta P$ to $Q$ 
\begin{equation}
 \Delta P  = \frac{\sigma^2}{k}\sqrt{Q}.\nonumber
\end{equation}
The square root  $\Delta P$ impact  is directly linked to the Kelly-optimal   $Q$ under the choice of portfolio diversification constraint\footnote{In this footnote it is shown that the same result can also be derived using the conventional terminology, where the leverage ratio $f$ is optimised instead of linking $Q$ directly with $\Delta P$.
To stay unit-consistent, we use the identity $P \, Q = f \, W$. We substitute the square root capital constraint, $W = k \sqrt{Q}$, to find the relationship between wealth $W$ and leverage $f$
$$W = k \sqrt{\frac{f W}{P}} \implies W = \frac{k^2 f}{P}.$$ 
Substituting $W$ back into the sizing identity: $$Q = \frac{f}{P} \left( \frac{k^2 f}{P} \right) = \frac{k^2 f^2}{P^2},$$ 
ergo leverage  $f$  scales quadratically with position size  $Q$. 
The instantaneous growth rate $g(f)$ is defined as the expected change in wealth minus the variance  $$g = \frac{E[dW]}{W} - \frac{1}{2} \frac{Var[dW]}{W^2}.$$ In an OU process with edge $\Delta P$ and volatility $\sigma$, the wealth change is driven by  $$dW = Q(\Delta P dt + \sigma dB).$$Substituting $Q = \frac{f W}{P}$ $$g(f) = \frac{f\,W\, \Delta P}{P\, W} - \frac{1}{2} \frac{(fW)^2 \sigma^2}{P^2 \,W^2},$$$$g(f) = \frac{\Delta P}{P} f - \frac{\sigma^2}{2 P^2} f^2,$$  and setting the derivative with respect to $f$ to zero leads to 
$$g'(f)=0=\frac{\Delta P}{P}-\frac{\sigma^2}{ P^2} f \,\,\Longrightarrow \,\,\Delta P \,P=\sigma^2 f,$$
  together with $$\sqrt{Q}\,P= fk,$$ results in
$$\Delta P=\frac{\sigma^2}{k} \sqrt{Q}.$$ 
Other choices for $f$, e.g.
 $f = const$, are inconsistent.  In an OU process, $\Delta P$ decays as the price moves toward the mean. If $f$ is constant, the growth rate $g  = f \Delta P - \frac{1}{2} f^2 \sigma^2$ would eventually become negative as $\Delta P$ becomes small. By using $f  \propto \Delta P$ the growth rate $g$ stays positive: $g  = \frac{1}{2} (\Delta P/\sigma)^2 \ge 0$. Trading stops exactly when the edge disappears, i.e. $\Delta P=0 \implies f=0$. 
 \\
What should be compared across assets is therefore not the growth rate, but the risk-adjusted edge, i.e. $f=\Delta P/\sigma^2$. 
The reason one  allocates the same amount  $f$  to two assets with different growth rates  $g$  is that capital allocation is a function of what could be termed  `leverage efficiency'. 
This can lead to paradoxes. For example, the increase of the investable universe can lead under special conditions to a drop in growth or even bankruptcy, see the Proebsting's paradox or the Braess paradox in the not completely unrelated field of traffic theory. 
}.

\noindent 
This stands in contrast to the single-asset portfolio, where time-dependence is explicitly captured by the subindex $t$. The investment strategy is governed by $f_t W_t = P_t Q_t$, and  the self-financing condition   is
\begin{equation}
 d W_t = Q_t d P_t.\nonumber
\end{equation}
The optimal fraction for the previously introduced OU process is
\begin{equation}
f_t= \frac{\kappa (P_t-K)}{\sigma^2}. \nonumber
\end{equation} 
Therefore,
\begin{equation}
Q_t= W_t \frac{\kappa (P_t-K)}{P_t \,\sigma^2}. \nonumber
\end{equation}
For convenience $K$ is set to zero, and this leads to
\begin{equation}
d \log W_t =  \frac{\kappa }{ \sigma^2} dP_t,  \nonumber
\end{equation}
with
\begin{equation}
W_t= W_0 e^{\frac{\kappa}{\sigma^2}(P_t-P_0)},
\nonumber
\end{equation}
and 
\begin{equation}
  Q_t= W_0 \frac{\kappa }{ \sigma^2}
  e^{\frac{\kappa}{\sigma^2}(P_t-P_0)}.\nonumber
\end{equation}

\noindent
The discussion is in the next section broadened to fractional Ornstein-Uhlenbeck (fOU) processes.
\section{Derivation of the Impact Law for Informative Order Flow}
\noindent 
The extension of the impact rule to general fOU processes,
\begin{equation} 
dP_t = \kappa  (P_t- K) dt + \sigma dB^H_t,
\nonumber
\end{equation}
with any Hurst exponent, $H \in (0,1)$, is carried out next.
For details about fOU see \cite{biagini2008stochastic}. The variance of the portfolio for a position size $Q$ invested in asset $P_t$ over time $T$ scales as $Q^2 \sigma^2 T^{2H}$. Under the portfolio capital constraint $W = k\sqrt{Q}$, the Kelly growth rate becomes,
\begin{equation}
g(Q) = \frac{Q T\Delta P}{k\sqrt{Q}} - \frac{\sigma^2 Q^2 T^{2H}}{2k^2 Q}. \nonumber
\end{equation}
Assuming the number of traded units  $Q$ scales linearly with $T$, i.e. $T \propto Q$ or $T = \hat{k}  Q$, the growth equation simplifies to
\begin{equation}
g(Q) = \frac{\Delta P}{k} \hat{k}  Q\sqrt{Q} - \frac{\sigma^2}{2k^2} \hat{k}^{2H} Q^{2H+1}. \nonumber
\end{equation}
Taking the derivative with respect to $Q$ and setting it to zero of $g(Q)/T$, since one is interested in the optimal growth per unit of time, one gets
\begin{equation}
 \frac{\Delta P \sqrt{Q}}{2k} - \frac{2H\,\,\sigma^2\,\, \hat{k}^{2H-1}}{2k^2} Q^{2H} = 0, \nonumber
\end{equation}
and 
\begin{equation}
\Delta P =   \frac{2H\, \hat{k}^{2H-1}\,\,\sigma^2}{k}  Q^{2H-\frac{1}{2}}\,\,. \nonumber
\end{equation}

\noindent 
An LP’s adjustment of the market impact function in the face of skewed order flow is not malicious; it is instead a natural defence against toxic flow and distinct from informational rent-seeking like front-running. In this context, `making a mistake is human, but repeating it is a choice'. Being `picked off' is a signal, and to neglect this information and leave the impact function unchanged would be an error. 

\noindent
The above formulation of market impact stands in contrast to the `linearised' liquidity models prevalent in decentralised exchanges. 
\section{Informational Attrition: CPMM eroded by Toxic Flow }
\noindent
In a standard CPMM, the invariant is $X \,Y = K$. The token exchange rate $P$ is the ratio of the reserves $P: = Y / X$. The associated market impact is derived next. Assume a LT buys a small amount $\Delta X$, the reserves change to $X + \Delta X$ and $Y - \Delta Y$. The new price $P_{new}$ is $$P_{new} = \frac{Y - \Delta Y}{X + \Delta X} = \frac{K}{(X + \Delta X)^2}.$$
A Taylor expansion around $\Delta X = 0$, $$P(X + \Delta X) \approx P(X) + P'(X) \, \Delta X,$$
combined with $P(X) = K X^{-2}$ and     $P'(X) = -2 k X^{-3} = -2 \frac{P}{X}$
leads to $$2\frac{ \Delta P}{P} \approx -  \frac{\Delta X}{X}  \, .$$
\noindent 
This is in conflict with the 
square-root and more generally Hurst dependent market impact found above.

 \noindent
\noindent
In an environment where block updates are orders of magnitude slower than high-frequency execution,  DEXs for many liquid assets   lag   CEXs, causing the local Hurst exponent to approach $H \approx 1$. This is especially acute, if the price on the CEX due to information inflow jumps.    
One is confronted with an `entropy leak' where the LP is under-compensated for the variance drag of  `toxic  flow'. While linearization   attracts volume, it creates a path-independent step-wise `crawl' that can subsidize  arbitrageurs.  The path-independence of CPMMs is in this case  a liability.
The problem is amplified in protocols that concentrate liquidity and heap on leverage.

\noindent
 As a simple consequence of the above result, the Kelly-optimal  market impact must be steeper than linear at the origin for $H<3/4$, i.e. sub-linear equates to $2H-1/2<1$.
Misjudging the Hurst exponent and implementing the wrong impact function does not necessarily lead to loss of capital, but can instead be associated with opportunity cost, especially for $H$ trending towards zero. Sensitivity analysis determines the amount of leakage due to misspecification.

\noindent
One way to ameliorate the cost to DEX LPs  is to penalize small orders and shift  execution prices closer to the post-execution pool ratio. The advantage and disadvantages of this and other proposals will be discussed elsewhere. 
For a fuller analysis one also has to consider `impermanence loss' and  `loss-versus-rebalancing' (LVR)\cite{milionis2022automated} for different proposals under a range of $H$.
  
\noindent
\noindent 
\noindent

\noindent




\noindent
Next follows sections on  the Carnot engine.
\section{A DEX trading strategy mapped into Thermodynamics}
\noindent
In this section the workings of a heat engine and a DEX trading strategy are positioned side by side after some introductory definitions.   

\begin{figure}[h]
 \includegraphics[width=10cm]{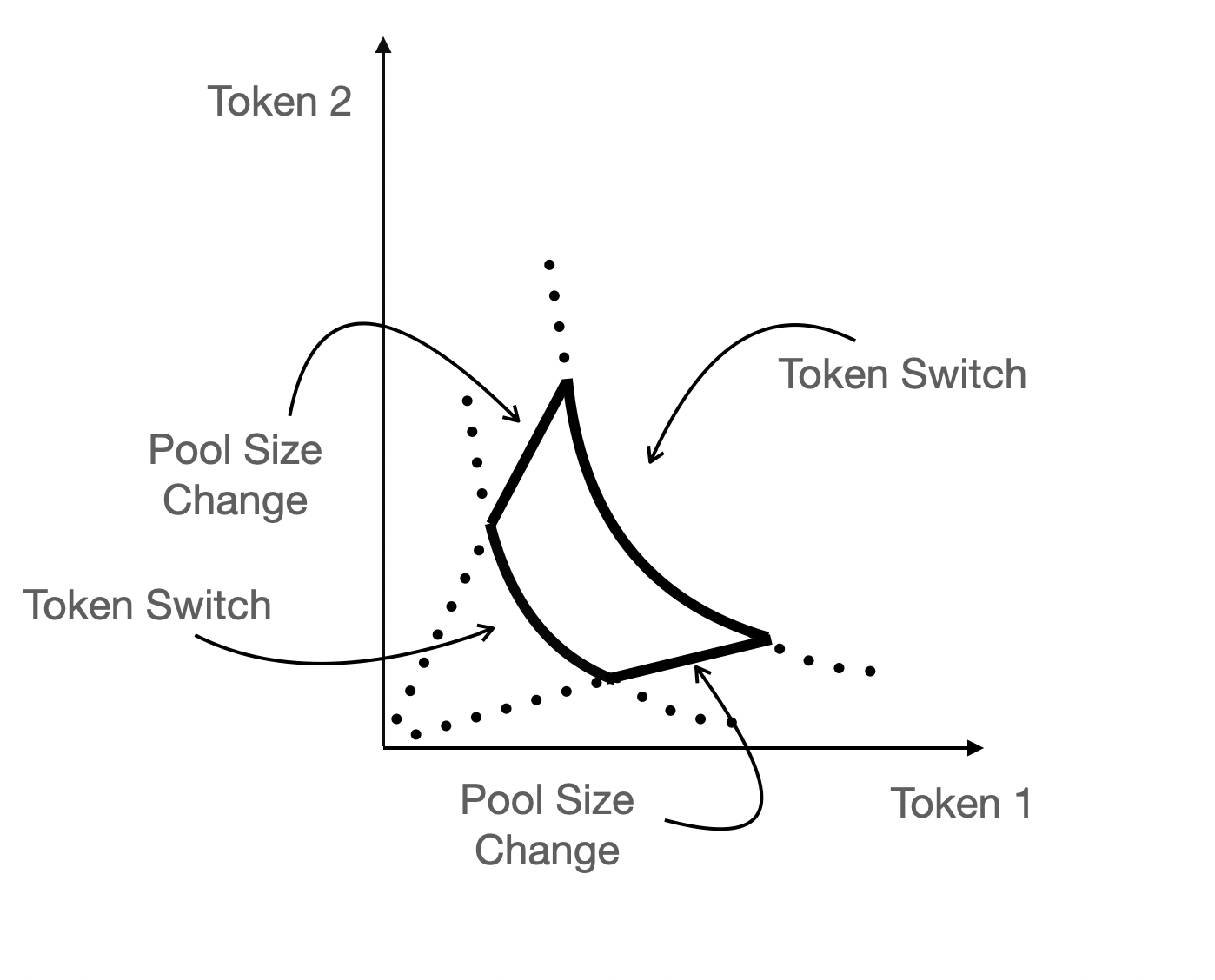}
\caption{ {\bf Crypto Carnot Cycle}: Four stage cyclic DEX trading strategy with two switching and two pool liquidity modification stages.} 
\end{figure}

\noindent
 Thermodynamics - Callen\cite{callen} provides  background - holds an exalted position in science. 
As Arthur Eddington famously remarked, `the law that entropy always increases holds, I think, the supreme position among the laws of Nature. If someone points out to you that your pet theory of the universe is in disagreement with Maxwell's equations - then so much the worse for Maxwell's equations. If it is found to be contradicted by observation - well, these experimentalists do bungle things sometimes. But if your theory is found to be against the Second Law of Thermodynamics I can give you no hope; there is nothing for it to collapse in deepest humiliation.'
 
 \noindent
 The ideal gas law, $PV=nkT$, finds a direct translation into the mechanics of  CPMMs. By mapping the reserve amounts $X$ and $Y$ to pressure $P$ and volume $V$ respectively,   $K$ emerges as the analogue to the system's thermal energy $nkT$,   where $n$  is the number of moles, $k$ is the ideal gas constant, and $T$ is the temperature. Within this framework, a token swap represents an `isoenergetic process', where the state variables $X$ and $Y$ adjust while the product $K$ remains invariant. Conversely, the provision or withdrawal of liquidity at the prevailing price ratio constitutes what could be termed an `isothermal process'\footnote{Energy is an extensive quantity, while temperature  is an intensive quantity, and the analogy between thermodynamics and DeFi has to be taken with a grain of salt. }.  The $X/Y$ ratio is then the system's temperature.
   \noindent
 The cyclic process\cite{bbm2000,bbm2002}  with four stages, see figure one, is next described in more detail.
 The initial and final amounts in the pool are for the first token $X$ and for the second token $Y$ with $XY=K$. At each stage an investor takes an action.
 In stage one and three an investor acts as a liquidity taker and switches tokens, whereas in stage two and four an investor expands and contracts the size of the pool.

 \noindent
\hspace{0cm}Side by side in the table below are the four stages of the cyclic processes: \\
\twocolumngrid
Thermodynamics \,\\ \,\\ \,
\,\,\,\,\,\,\,\,\,\,\,\,DeFi 
\onecolumngrid

\vspace{.1cm}
\hspace{6.1cm}\underline{Step one}
\vspace{.1cm}
\twocolumngrid
Isothermal Expansion\\ \,\\ \,\\  \,\\
Switching $\alpha$ of the first token to $\beta$ of the second token while leaving $K$ unchanged leading to $X-\alpha $ and $Y+\beta$ with $(X-\alpha)(Y+\beta) $  
\onecolumngrid
\vspace{.21cm}
\hspace{6.1cm}\underline{Step two}
\vspace{.1cm}
\twocolumngrid
 Isentropic Expansion
\\ \,\\ \,\,\\
 Adding $M$ of the first and $N$ of the second token in the right ratio, i.e. while $(X-\alpha)/(Y+\beta)=M/N $ 
\onecolumngrid
\vspace{.21cm}
\hspace{6.1cm}\underline{Step three}
\vspace{.1cm}
\twocolumngrid
Isothermal Compression
\\ \,\\ \, \\ \, \\
Switching $\delta$ of the second token to $\sigma$ of the first token with  
$(X-\alpha+M)(Y+\beta+N)=(X-\alpha+M+\sigma)(Y+\beta+N-\delta)$
\onecolumngrid
\vspace{.21cm}
\hspace{6.1cm}\underline{Step four}
\vspace{.1cm}
\twocolumngrid
Isentropic Compression
\\ \,\\ \,\,\\
Subtracting $G$ of the first and $H$ of the second token in the right ratio, i.e.
while $(X-\alpha+M+\sigma)/(Y+\beta+N-\delta)= G/H$
\onecolumngrid
\vspace{.21cm}
 %
\noindent
In thermodynamics the  area enclosed corresponds to the work, and in DeFi the net gain or loss should have a similar interpretation.

\noindent
We assume the operator of the engine, as well as the trader accessing the DEX, can carry out the four stages without interference, i.e. there is complete control by the trader over the incorporations of transactions into the blockchain.
The following caveats further apply.
A `perfect' engine is considered with no extraneous losses, and in the trading context, a
`perfect' market with neither transaction or nor gas cost.
The next paragraphs provide an example calculation and implicitly answer the question.
What takes the role ofheat baths in CPMM? In thermodynamics the Carnot efficiency bound for a heat engine with access to two heat baths is given by $1-T_1/T_2$ with $T_2\geq T_1$.


\noindent
The calculation accompanying the four stages of the crypto Carnot cycle are given next. the starting and final composition of the pool will be identical. 
There are three parts to the description of the token composition at each stage. These are the total token amounts in the pool, the token amounts owned by the investor in the pool, and the   token amounts owned by the investor outside the pool.  Temporary short position are allowed. This could involve a marginal amount of borrowing cost, which is ignored. \\\\\\\\

\vspace{2.012cm}
\underline{Starting point:}\\
Pool: (X,Y) with XY=K, i.e. the pool composition is a two component vector of amount $X$ of the first and  amount $Y$ of the second token.\\
Investor: \\
Investor position inside pool : $(0,0)$\\
Investor position outside pool: $(0,0)$\\

\underline{Result of stage 1:} Switch of $\alpha$ of the first token for $\beta$ of the second token. \\
Pool: $(X-\alpha,YX/(X-\alpha))$\\
Investor: \\
Investor position inside pool : $(0,0)$\\
Investor position outside pool: $(\alpha,-\beta)$\\
with $\beta=XY/(X-\alpha)-Y$\\

\underline{Result of stage 2:} Addition of $M$ of the first token and $N$ of the second token. \\
Pool: $(X-\alpha+M,XY/(X-\alpha)+N)$\\
Investor: \\
Investor position inside pool : $(M,N)$\\
Investor position outside pool: $(\alpha-M,-\beta-N)$
with $N=M\,Y\,X/(X-\alpha)^2$\\

\underline{Result of stage 3:} Switch of $\delta$ of the second token for $\sigma$ of the first token. \\
Pool:$(X-\alpha+\sigma+M,YX(1+M/(X-\alpha))/(X-\alpha)-\delta)$\\
Investor:\\
Investor position inside pool : $(M,N)$\\
Investor position outside pool: $(\alpha-M-\sigma,-\beta-N+\delta)$\\
with $\delta=\sigma \frac{YX}{X-\alpha} \Big(1+\frac{M}{X-\alpha}\Big)\frac{1}{X+M}$\\

\underline{Result of stage 4:} Removal of $G$ of the first token and $H$ of the second token. \\
Pool: $(X-\alpha+\sigma+M-G, \frac{YX}{X-\alpha}(1+\frac{M}{X-\alpha})-\delta-H)$. \\ 
Investor:\\
Investor position inside pool : $(M-G,N-H)$\\
Investor position outside pool: $(\alpha-M-\sigma+G,-\beta-N+\delta+H).$ \\
The four stage cycle does not generally close, and even if by setting $M-G-\alpha+\sigma=0$ and $\frac{YX}{X-\alpha}\Big(1+\frac{M}{X-\alpha}\Big)-\delta-H=Y$ the 
final pool composition is again $(X,Y)$, the final investor position inside and outside the pool will not have collapsed back to zero. 


\noindent
The strategy described above is not capital free\footnote{ A requirement exists to borrow tokens temporarily to add to the pool in stage two. This needs capital.  The same holds true for stage one and three to enable the investor to switch tokens. This leads to borrowing cost and collateral requirements and adds a twist to the analysis ignored in the idealised setting. Flashloans can potentially reduce  the capital employed.}. Friction  associated with pool dependent transaction fees,   variable gas fees and the need to access the blockchain without for example MEV interference was ignored,  complicates the situation and weakens any obtainable profit. 
Viewed geometrically, transaction costs distort the crypto-equivalent `isothermal curves'. A formal investigation reducing these frictional effects to a system of inequalities will be carried out elsewhere.
Stablecoin de-pegging risk is the topic of the next part.
\section{ Kelly and the Cat-bonds:  Optimality in Catastrophe Bond Portfolios}
\noindent
How   stablecoin de-pegging, both temporary as well as irreversibly,  can be covered by catastrophe insurance   either in form of conventional cat-bonds or on-chain is investigated.   The passing of  `The Guiding and Establishing National Innovation for U.S. Stablecoins Act' (GENIUS Act) has spawned a multibillion-dollar yield windfall by forbidding issuers to pay interest to stablecoin holders. Foregone interest collected by issuers could be partially pumped into insurance premiums, subsidizing the cost of a private-market backstop.  This has been an active topic of discussion, but the market remains nascent, held back by legal thickets and wrangling.

\noindent
In a recent paper\cite{eichengreen2025} estimates were given for   de-pegging risk. 
It will be shown how to relate the price of different catastrophe risks, e.g.  de-pegging priced on the back of catastrophe bond benchmarks.


\noindent 
Stablecoins  - whether backed by a vault or algorithmic - have suffered  crashes. They are still a work in progress, even as hundreds of billions of dollars courses through the system. While they aim to be as steady as cash, their history is dotted with sudden melt-downs, rendering them into a risk to be managed.



 \noindent

 
 \noindent
 
\noindent
The Kelly criterion can provide this relative pricing of catastrophe bonds.
Two different basic edge cases are considered. 
First, either a bond $A$ or bond $B$ is contained in the portfolio. In the second method, both bond $A \& B$ are held in the same portfolio and the relative weights are determined. 

\noindent
Discrete and continuous price processes lead to different Kelly optimizations. 
In the one dimensional continuous case the optimal fraction is under relatively general conditions  given by the ratio of the excess return minus the short rate and variance. In the multidimensional case the excess return is a vector multiplied by the inverse (or more generally the Moore-Penrose inverse\cite{meister2011}) of the covariance matrix. For independent assets it would be a diagonal matrix. This restriction to the first and second moment does not hold when one considers optimization for discrete payouts, where higher order moments cannot be neglected. The discrete view is more natural for catastrophe payouts and will be used here, nevertheless even in the discrete case the ratio of the mean and variance gives a first estimate for the optimal fraction. It provides the simplest way to relate different catastrophe risks.  

\noindent
If the portfolio has only access to one  risky asset with $q$ the default probability, $p=1-q$ the corresponding survival probability, $f$ the fraction invested in the risky assets, and $r$ the return in the non-default case. Recovery in the case of a default is set to zero.   The growth expression to maximize is
 \begin{eqnarray}
q\log(1-f)+p\log(1+fr)
\nonumber 
\end{eqnarray}
to get 
 \begin{eqnarray}
f=1  - q - \frac{q}{r},\nonumber 
\end{eqnarray}
which corresponds to the conventional statement of the formula in the double-or-nothing betting game for $r$ equal to one.
How can the optimal $f$ remain unchanged, if  $q$ is replaced by $q+\Delta$ and $r$ by $r+\delta$? This requires the following relationship between the additive  adjustments
 \begin{eqnarray}
\epsilon& \approx& \frac{\delta}{r} \left( \frac{q}{1 + r} \right),\nonumber\\
& \approx& \frac{\delta}{r} \left( \frac{q}{1+r} + \frac{q^2}{(1+r)^2} + \frac{q^3}{(1+r)^3} + \dots \right).
\nonumber 
\end{eqnarray}
\noindent


 \noindent

 \noindent
As a next and final step, we consider the two bond portfolio  with $f_A=f_B:=f^*$ and include the lowest cross term, but instead assume $q_A=q_B:=q$ and therefore $r_A=r_B:=r$ to get
$$f^* \approx \frac{1}{2} - \frac{q}{r} - \frac{q^2}{2r^2} -\frac{q}{3r} \left( \frac{q}{r} \right)^2,$$
where the last term is a cross term capturing that both bonds can default simultaneously.
To reiterate,  investment fraction equality does not translate into growth equality, especially when rare events are involved, and as already mentioned at the end of footnote 2 this can lead to paradoxes.   

 \noindent
 The two toy examples explored  how in idealized settings  catastrophe risks can be related.
 Similar and more interesting scenarios have been considered   by Ziemba and others.




\noindent

\noindent

\noindent


\noindent

\noindent


\noindent

\noindent

\section{Conclusion}
\noindent
Derivations based on information theory,  called colloquially `Kelly optimisation' in some finance applications\cite{bkm2016, meister2022, meister2023, meister2024, meister2025}, can   provide insights into market-place design conundrums. 

\noindent
After deriving market impact for a range of Hurst exponents, it was shown how this differs from the artificial market impact implemented on CPMMs.
As $H$ moves towards extremes the  capital misallocations of LPs on CPMMs are exacerbated, and the LPs liquidity provisions becomes decoupled from the true growth-optimal pricing curve, leading to both sub-optimal fee accumulation and increased exposure to uncompensated jump-risk.  
 
\noindent
Market impact arises from the bottom-up competition inside individual LPs for capital allocation. However there is also top-down competition between LPs enforcing price consistency across the market - the `law of one price' - even as the $Q-\sqrt{Q}$ effect  grants advantages that widen with scale. This is balanced by drawbacks, since a `monopoly' LP becomes the market itself and, as `a whale in a tub', is unable to stir  due to liquidity constraints. The critical threshold is easier to observe in a crisis, as one saw with Archegos Capital Management or earlier with LTCM, then to predict beforehand, for as Keats noted, `axioms in philosophy are not axioms until they are proved upon our pulses'.

\noindent
Tail risk associated with various de-pegging events hampers the influx of institutional and retail money to decentralized finance. 
Enumerating these risks and providing hedging tools through catastrophe bonds or similar instruments enables broader institutional and   retail adoption. It would expand the range of uncorrelated investments available to tail risk funds on the demand side and stable coin issuers could use it on the supply side to distinguish their coins from competitors.  The availability of protection can enhance the attractiveness of a stable coin and can be viewed as a cost efficient marketing, especially when due to the GENIUS Act   payment of interest on stable coin deposits is forbidden.

\noindent
De-pegging risk   is a tuneable risk similar to an option strike, since it ranges all the way from a temporary deviation of a chosen percentage from the peg to an irreversible failure. This gives design flexibility to obtain  yields ranging from  single to low double digits. 




\noindent



\noindent

\noindent


\noindent
Paul Volcker quipped after the 2008 financial crisis that `ATMs were the only useful innovation in banking for the past 20 years'.  One might say the same for stablecoins. They are the plainest part of DeFi, but they fill a market need. In doing so, they take on the main role of fiat money  as a means of exchange. Other assets can handle the other two  functions: being a store of wealth and a unit of account.


\noindent
Tweedledee spells out the looking-glass logic\footnote{W. H. Auden’s 1962 essay, ``Today’s ‘Wonder-World’ Needs Alice'', draws a parallel between a rational Alice caught in a   maze of upside-down logic and a modern world of vanishing certainties.
  } of DeFi: 'Contrariwise, if it was so, it might be; and if it were so, it would be; but as it isn't, it ain't. That's logic.'   

 
 \end{document}